\documentclass[12pt,preprint]{aastex} 

\usepackage{graphicx}

\received{}
\accepted{}

\slugcomment{To be submmitted to ApJ Letters}
\lefthead{Wilkes et al.}
\righthead{X-ray properties of Red Active Galactic Nuclei.}
\newcommand{\etal}{{\it et~al.}}
\newcommand{\alpe}{$\alpha_E$}
\newcommand{\nh}{${\rm N_H}$}
\newcommand{\lax}{${_<\atop^{\sim}}$}
\newcommand{\gax}{${_>\atop^{\sim}}$}

\begin{document}

\title{The X-ray properties of 2MASS Red Active Galactic Nuclei}

\author{Belinda J. Wilkes\altaffilmark{1}, Gary D. Schmidt\altaffilmark{3}, Roc
M. Cutri\altaffilmark{2}, Himel Ghosh\altaffilmark{1}, Dean C. Hines\altaffilmark{3},
Brant Nelson\altaffilmark{2}, \& Paul S. Smith\altaffilmark{3}}

\altaffiltext{1}{Harvard-Smithsonian Center for Astrophysics, Cambridge, MA 02138}
\email{bwilkes,hghosh@cfa.harvard.edu}

\altaffiltext{2}{IPAC, Caltech, MS 100-22, Pasadena, CA 91125}
\email{roc,nelson@ipac.caltech.edu}

\altaffiltext{3}{Steward Observatory, University of Arizona, Tucson, AZ 85721}
\email{dhines,gschmidt,psmith@as.arizona.edu}

\shorttitle{X-RAY PROPERTIES OF RED AGN}
\shortauthors{WILKES, SCHMIDT, CUTRI, GHOSH, HINES, NELSON, SMITH}

\begin{abstract}

The Two Micron All Sky Survey (2MASS) is finding previously unidentified,
luminous red active galactic nuclei (AGN).  This new sample has a space density
similar to, or greater than, previously known AGN, suggesting that a large
fraction of the overall population has been missed. {\it Chandra\/}
observations of a well-defined subset of these objects reveal that all are
X-ray faint, with the reddest sources being the faintest in X-rays.  The X-ray
hardness ratios cover a wide range, generally indicating \nh $\sim 10^{21-23}$
cm$^{-2}$, but the softest sources show no spectral
evidence for intrinsic absorption.
These characteristics suggest that a mix of absorbed, direct emission and
unabsorbed, scattered and/or extended emission contributes to the X-ray flux,
although we cannot rule out the possibility that they are intrinsically X-ray
weak.  
This population of X-ray faint, predominantly broad-line objects
could provide the missing population of X-ray absorbed AGN required by current
models of the cosmic X-ray background.  The existence of AGN which
display both broad emission lines and absorbed X-rays has important
implications for unification schemes and emphasizes the need for care in
assigning classifications to individual AGN.

\end{abstract}


\section{Introduction}

The realization that obscuration plays a critical role in the classification of
AGN inspired a fundamental change in our understanding of the phenomenon.  Not
only does the ``Unified Scheme'', in which narrow emission line (type 2) AGN are
interpreted as edge-on broad emission line (type 1)
AGN, provide a basis for new observations and
theoretical models, we also realize that many AGN may be nearly invisible in
UV-excess surveys ({\it e.g.,} Webster \etal\ 1995; Masci \etal\ 1999).  This
expanded idea of what comprises an AGN means that their current number density
may be significantly underestimated ({\it e.g.,} Sanders and Mirabel 1996).  Ramifications include revisions of the fraction and types of galaxies that
harbor an active nucleus, the energy density of ionizing flux in the young
universe, and the nature of the X-ray and far-IR backgrounds.

Although {\it IRAS\/} provided the first significant sample of extragalactic
objects in which the bulk of the luminosity emerges as reprocessed radiation in
the IR (Soifer \etal\ 1984), its sensitivity was sufficient to catalog only the
most nearby and/or luminous AGN. The Two Micron All Sky Survey (2MASS) is
yielding a much deeper catalog of near IR-selected AGN (Cutri \etal\ 2001) by
selecting sources with $J-K_s > 2$ from the high galactic latitude 2MASS Point
Source Catalog.  Spectroscopic follow-up of red candidates reveals $\sim$75\%
are {\it previously-unidentified} emission-line AGN, with $\sim$80\% of these
showing broad optical emission lines (Type 1: Seyfert 1 and QSO), and the
remainder being narrow-line objects (Type 2: Seyfert 2, QSO 2, and LINER; Cutri
\etal\ 2001).  They span a redshift range $0.1<z<2.3$ with median $\sim$0.25.
The inferred surface density is $\sim$0.5~deg$^{-2}$ brighter than
$K_s=14.5$ mag., higher than that of
optically selected AGN at the same IR magnitudes and indicating that 2MASS will
reveal $>$25,000 such objects over the sky.  The objects
have unusually high optical polarization levels, with $\sim$10\% showing
$P>3$\% indicating a significant contribution from scattered
light (Smith \etal\ 2001).

{\it ROSAT\/} found that, while known AGN dominate the soft ($0.1-2.0$~keV)
cosmic X-ray background (CXRB; Lehmann \etal\ 2000), an additional population
of heavily absorbed AGN would be required to account for the harder high-energy
spectrum (Comastri \etal\ 1995).  To match both the CXRB spectrum and the
observed hard X-ray number counts of pre-{\it Chandra\/} surveys (Fiore \etal\ 2001),
the X-ray absorbed AGN population is estimated to outnumber unabsorbed AGN by
$\sim$4:1 and perhaps to increase with $z$ (Gilli \etal\ 2001; Comastri
\etal\ 2001). Although the ratio of Type 2 to Type 1 AGN in the local universe
is consistent
with this: $\sim$$2-4$ (Maiolino \& Rieke 1995; Huchra \& Burg 1992),
a dominant population of X-ray absorbed AGN at z\gax 0.1 has yet to
be found.
Possible identifications include ADAF (advection dominated accretion flow)
galaxies (Di Matteo \etal\ 1999) and
narrow emission-line X-ray galaxies with flat/absorbed X-ray spectra, luminous
IR galaxies (Risaliti \etal\ 2000), and a subset of broad-line AGN that are
X-ray absorbed.

The {\it Chandra\/} X-ray Observatory (Weisskopf \etal\ 2000)
and {\it XMM-Newton\/}
(Jansen \etal\ 2001), with their faint flux limits and broad energy sensitivity
($\sim$$0.5-10$ keV),
are finding objects in sufficient numbers to explain $60-80$\% of the CXRB,
including a significant number of hard spectrum sources.
These correspond to both optically faint objects and bright, nearby, but
otherwise normal elliptical galaxies (Hornschemeir \etal\ 2000; Barger
\etal\ 2001; Giacconi \etal\ 2001), as well as more traditional, broad-line
AGN.  Near-IR observations of the latter reveal featureless, red
continua ($1.5 < J-K_S < 2.5$) which, combined with their optical colors, are
consistent with the moderate amounts of absorption required to match
the CXRB (Compton-thin; equivalent neutral hydrogen column density $N_{\rm
H}~<~10^{24}$~cm$^{-2}$; Crawford \etal\ 2001; Barger \etal\ 2001).  Thus,
evidence is mounting that absorbed AGN are indeed important contributors to the
CXRB.

The area covered by the deepest {\it Chandra\/} surveys is small ($\sim$0.16
sq. deg.), resulting in large statistical uncertainties, particularly at low
$z$.
Given the high surface density and similarities of the new population of 2MASS
AGN to the smaller {\it Chandra\/} sample, a census of the X-ray properties of
the 2MASS AGN will likely be an essential ingredient in their understanding and
will yield an estimate of whether this previously missed population is
sufficient, alone, to explain the shortfall in the CXRB.

\section{Observations}

We are surveying a well-defined, flux-limited, and color-selected subset of 26
2MASS AGN using the Advanced CCD Imaging Spectrometer array (ACIS-I; Nousek
\etal\ 1998) on {\it Chandra\/}.  The subset was selected
to have $B-K_S>4.3$ and $K_S<13.8$, including the brightest and reddest objects
but covering sufficient parameter space to be representative of the new
population:  $0<z<0.4$, $0<P(\%)<9.3$ (Smith \etal\ 2001).
Included are six narrow-line (Type 2) AGN and 20 with broad optical emission
line components (Type 1, 1.5, 1.8).  Because the observed 
$K_S$-band to X-ray slopes for AGN span a wide range,\footnote{Sy2s
and Broad Absorption
Line (BAL) QSOs lie at the extreme X-ray faint end of this range.}
(1.1\lax $\alpha_{KX}$\lax 2 Elvis
\etal\ 1994; Lawrence \etal\ 1997), the X-ray flux expected from the 2MASS
sample is uncertain by a factor $\sim$800, even if they have properties
similar to ``normal" AGN.
Observing times ranging from $1-4.5$ ksec were selected to acquire at least 5
counts at the faintest end of this range.

X-ray counts and hardness ratios for the 23 targets observed to date 
are summarized in Table~\ref{table}. Virtually
all source counts are contained in the 2.9\arcsec\ (5.75 pixel) radius
extraction aperture that was centered upon the source.  Background counts were
estimated from an annulus with inner and outer radii of 10 and 30 pixels,
respectively.  Four targets are undetected.

\section{Spectral Energy Distributions}

\begin{figure}
\plotone{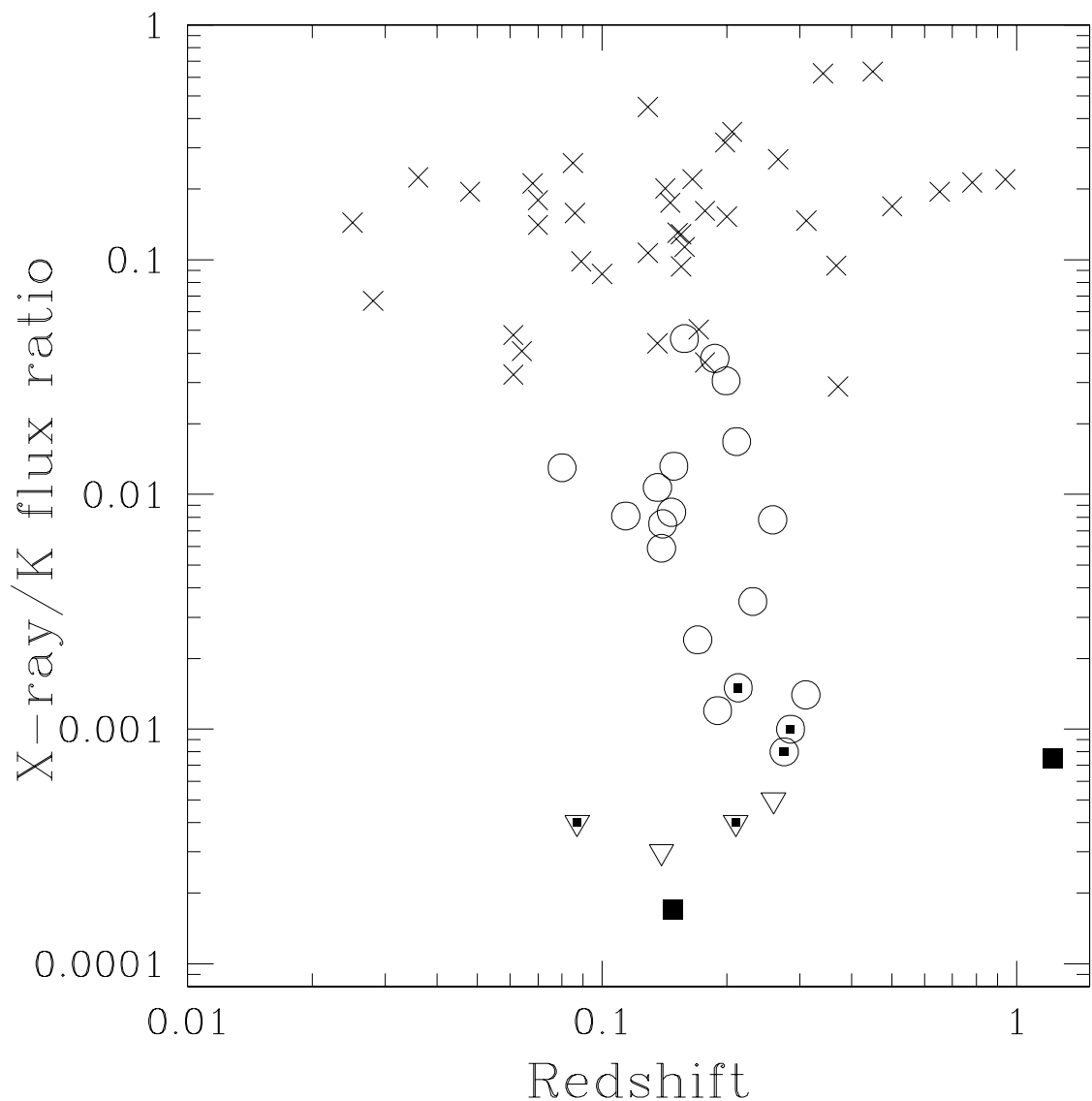}
\caption{The observed X-ray-to-$K_S$ flux ratio (1 keV X-ray flux assuming a
normal AGN X-ray spectrum: \alpe = 1.0, \nh\ = 3$\times 10^{20}$ cm$^{-2}$) as a
function of redshift for the 2MASS AGN (circles) compared with the same ratio
for low-redshift, broad-line AGN (Elvis \etal\ 1994; crosses).  Upper limits
are indicated by triangles and the reddest sources ($J-K_S>2.5$)
by a central dot. The squares show the BAL QSOs IRAS07598+6508 and FIRSTJ0840+3663
(Green \etal\ 2002, Low \etal\ 1989, 2MASS survey).
\label{fg:redX}}
\end{figure}

A comparison of the $K_S$-band to 1 keV flux density ratios (the latter
computed assuming a ``normal'' AGN X-ray spectrum; \nh=$3\times 10^{20}$
cm$^{-1}$, \alpe=1.0, where F$_{\nu} \propto \nu^{-\alpha_E}$) with those of
low-redshift, broad-line AGN (Figure~\ref{fg:redX}) demonstrates the general
weakness of X-ray emission from the 2MASS QSOs, placing them in the range
measured for BAL QSOs and Sy2 galaxies,  Figure~\ref{fg:redX}).
We also find that the objects with the
reddest colors ($J-K_S>2.5$) are the weakest X-ray sources ($>$99\%
significance, Figure~\ref{fg:redX}).
This suggests either an underlying, non-thermal power law with a range
of slopes dominating the near-IR and X-ray spectral regions
(Carleton \etal\ 1987, Brissenden 1989) or correlated absorption 
in the near-IR and X-ray regions.
The latter would predict hard X-ray spectra for these sources.

\begin{figure}
\includegraphics{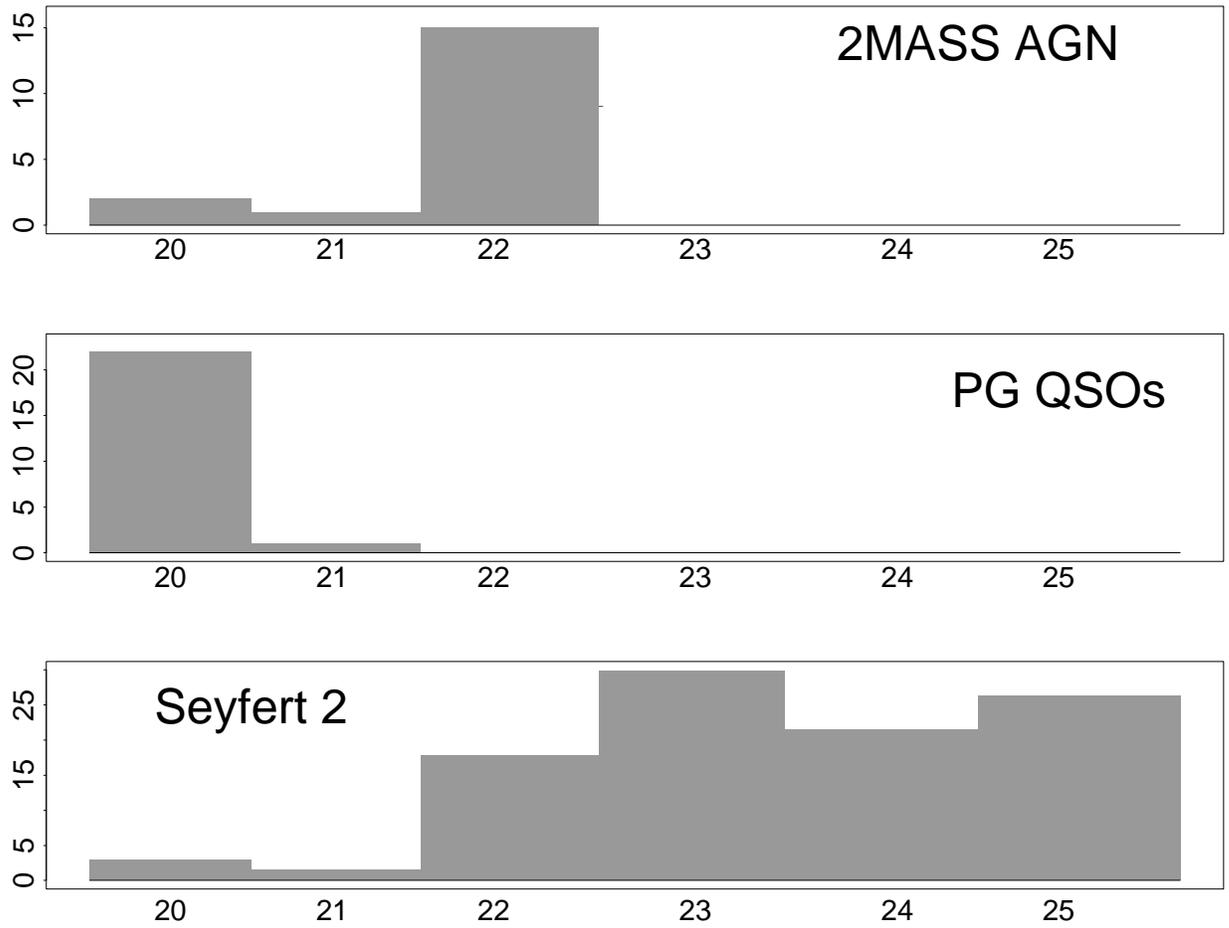} 
\vskip 1.0in
\caption{The distribution of the logarithm of the equivalent neutral hydrogren
absorption column densities, \nh, derived from the X-ray hardness ratios for
the 2MASS AGN assuming a power law spectrum with \alpe=1.0 and absorption
intrinsic to the AGN, compared with those of optically selected PG QSOs (Laor
\etal\ 1997) and Seyfert 2 galaxies (Risaliti \etal\ 1999).
\label{fg:nhdist}}
\end{figure}

The low signal-to-noise (S/N) ratio of most of the observations does not permit
direct measurements of the X-ray spectral slope and absorption.  Instead, we
evaluate the X-ray hardness ratio (Table~\ref{table}).
A normal AGN X-ray
spectrum with little absorption (\alpe\ $\sim$~1.0) yields a {\it Chandra\/}
hardness ratio of $-$0.7 in the bands used here.  The 2MASS sample covers the
range $-0.6 <$ HR $< +0.6$, consistent with \nh\ $\sim 10^{21-23}$ cm$^{-2}$
for most sources.  One 2MASS AGN of Type 1.5 and one of Type 1 are consistent
with no absorption above the Galactic column (HR $<-0.5$).  
These results are
shown in Figure~\ref{fg:nhdist}, where it is seen that the 2MASS sources tend
to fall intermediate between values derived for the optically selected PG QSOS
(Laor \etal\ 1997) and the Seyfert 2 galaxies (Risaliti, Maiolino, \& Salvati
1999).

There is no
relation between HR and $J-K_S$, {\it i.e.} while X-ray fainter, the reddest
objects do not have the hardest spectra, as might be expected in a
simple optical and X-ray absorption scenario.  Neither
X-ray flux nor HR appears to be related to AGN class or to optical polarization
level.
While these estimates of \nh\ are instructive, 
hardness ratios are, at best, a crude measure of the true \nh.
The X-ray spectra of AGN often contain a number of components
which modify the power law continuum emission believed to originate
in the central AGN. In addition to the cold absorbing material we assumed
above, these include partial covering of the
source, an ionized absorber, extended emission, and scattering/reflection
by neutral and/or ionized material.
For example, the spectrum of NGC~1068 is a mix of neutral and ionized
reflection of an otherwise invisible nuclear continuum (Matt \etal\ 1997)
so that the observed hardness ratio is de-coupled from the intrinsic
cold absorber.
Without spectral deconvolution of high S/N data, it is not possible to
distinguish between the various possibilities.  Correction of the X-ray flux
for absorption as suggested by the hardness ratios moves only about half of
the 2MASS sample
into the area occupied by the Elvis \etal\ (1994) AGN in Figure~\ref{fg:redX},
suggesting that complex spectra result in excess soft emission in many sources.
However,
we cannot rule out the alternate possibility that the X-ray emission of at
least some 2MASS AGN is dominated by a power law that is intrinsically faint
and hard.

\section{Optical and X-ray Absorbing Material}

Assuming that dust extinction is responsible for the red colors of 2MASS AGN
and absorption by associated gas for the hardness of the X-ray spectra, we can
compare the equivalent column densities on a case-by-case basis.  We assume
that the median, rest-frame $J-K_S$ for an AGN is 2.04 (Elvis \etal\ 1994) and determine
$E(J-K_S)$ from the observed $J-K_S$ color. Comparison between this and
the X-ray-derived \nh\ shows that the ratio
E(J-K$_s$)/\nh\ is reduced by a factor of a few to $\sim$100 compared to the
Galactic value of $0.98 \times 10^{-22}$ cm$^2$ mag., the latter being
computed from the ROSAT dust extinction
vs. gas absorption relation (Seward 1999) and the extinction curve from Mathis (1999).
A similar result was reported for optical 
(E(B-V)) vs X-ray extinction in AGN by Maiolino
\etal\ (2001), who suggest an explanation in terms of anomalous dust --
dominated by large grains -- which absorbs with little spectral reddening.
Other possibilities include different lines of sight to the X-ray and IR
continuum emitting regions such that much of the X-ray absorbing material does
not cover the optical/IR source (Risaliti \etal\ 2000), or absorbing material
that is hotter than the dust sublimation temperature and so is largely
dust-free and/or physically de-coupled from the dust
(Granato, Danese, \& Franceschini 1997; Risaliti, Elvis, \& Nicastro
2001). Optical spectropolarimetry to study the scattering material
combined with improved constraints on the X-ray absorbing column density
for a significant number of these AGN will allow us to distinguish between
these various possibilities. 

A similar comparison of the reddening indicated by $B-K_S$ color with that from
the near-IR shows good agreement in most cases, particularly given the large
uncertainties associated with the B  magnitudes from the United States
Naval Observatory (USNO) A-2.0  Catalog (Monet
\etal\ 1998).
A small subset of the 2MASS AGN show less optical than near-IR reddening
probably due to 
host galaxy and/or scattered nuclear light contributing to the optical flux.
Although there is no clear correlation between the discrepancy and optical
polarization level, the source with the largest discrepancy ($\sim$3.7 mag.) has
$P=6$\%.
Alternatively additional IR emission from hot dust could
increase the light at $K_S$.

\section{Red AGN and the CXRB}

Current CXRB models require a population of X-ray absorbed AGN that has yet to
be found beyond the local universe. It is not essential that these objects be
classified as AGN at other wavelengths, but the various source populations
considered to date, including luminous IR galaxies, Seyfert 2s, QSO 2s, LINERs,
normal galaxies etc., are insufficient to account for the measured surface
brightness.

The new population of predominantly Type 1 AGN being discovered in
the 2MASS AGN survey at least doubles the density of broad-line objects at $z$
\lax\ 0.5.  The bias toward low-$z$ objects inherent in the 2MASS
color selection process implies that such a population is also present
at high $z$. If so, their numbers combined with the hardness of their X-ray
spectra imply that they can account for most of the missing CXRB population,
as opposed to Type 2 AGN or other heavily obscured objects.
However their high $z$ counterparts will need to be found by other means
such as {\it e.g.} looking for 
red sources in the longer wavelength SIRTF bands.
More distant absorbed, Type 1 AGN have been found, but they are not common
and are usually identified as
radio-loud quasars (Reeves \& Turner 2000, Gregg \etal\ 2001).

\section{The Nature of X-ray Absorbed, Broad-line AGN}

The presence in this sample of Type 1 AGN with absorbed X-ray spectra suggests
that a simple unification scenario in which edge-on AGN exhibit narrow optical
lines with the broad lines and X-ray emission being strongly absorbed, while
face-on counterparts are unobscured in both emission lines and X-rays, is too
simplistic. It also emphasizes that classifications may not apply beyond the
waveband in which they were made.

The combination of X-ray absorption, red near-IR continuum, polarized
optical continuum, and
broad lines in the majority of 2MASS AGN suggests that they are viewed at
an intermediate line of sight with respect to dusty, nuclear material
({\it e.g.} torus/disk/wind)
such as has been proposed for the similarly
polarized BAL QSOs. The X-ray absorbing medium must be mostly dust free,
perhaps being largely inside the sublimation radius, and patchy,
resulting in partial covering of the broad emission-line region. The emission
lines and/or optical continuum may include a scattered component, perhaps from
absorbing material on the far-side of the engine as has been suggested for BAL
QSOs (Schmidt \& Hines 1999; Ogle 1999; and references therein).  It seems
likely that a scenario similar to the accretion disk plus outflowing wind
proffered to explain various aspects of AGN ({\it e.g.} Weymann \etal\ 1991;
Voit \etal\ 1993; Murray \& Chiang 1995; Ogle 1999; Elvis 2000) could be
invoked for these objects as well. If so,
it is likely that many of the 2MASS AGN will show broad absorption lines
in their UV spectra.
Their X-ray weakness, along with that of optical Type 2s
and BAL QSOs also suggests that highly obscured objects may go
undetected even in the X-rays
so that weak or non-detected X-ray emission
does not rule out the presence of an active nucleus.
However, a more detailed study of their
optical and X-ray emission is needed before applying such a scenario in detail.


\acknowledgments

Fruitful discussions with Drs. Martin Elvis and Guido Risaliti are gratefully
acknowledged.  This publication makes use of data products from the Two Micron
All Sky Survey, which is a joint project of the Univ. of Massachusetts and the
Infrared Processing and Analysis Center/California Institute of Technology,
funded by NASA and the National Science Foundation. Financial support was
provided by NASA grant GO1-2112A (Chandra GO) and NASA contract NAS8-39073
(Chandra X-ray Center).
RMC and BN acknowledge the support of the Jet Propulsion Laboratory
which is operated by the California Institute of Technology
under contract to NASA.

\clearpage
\begin{deluxetable}{llrrrr}
\tablenum{1}
\tablewidth{0pt}
\tablecaption{{\it Chandra\/} Observations of 2MASS Red AGN \label{table}}
\label{table}
\tablehead{\colhead{Name} & \colhead{Date} & \colhead{Exp.} & \colhead{Cts.\tablenotemark{a}} & \colhead{HR\tablenotemark{b}}  & \colhead{J$-$K$_S$}
\\ \colhead{2MASSI}  & YMD & \colhead{(s)} & & & }
\startdata
000703+1554 &   010116 &   3533 &   99  &   0.5$\pm$0.1     &   2.129 \\ 
005055+2933 &   010125 &   3644 &   120 &   0.1$\pm$0.1     &   2.114 \\ 
010835+2148 &   010126 &   3100 &   10  &   0.2$\pm$0.4     &   2.754 \\ 
012031+2003 &   010126 &   1538 &   $<5$&   $\cdots$        &   3.851 \\ 
015721+1712 &   010126 &   2700 &   135 &   $-$0.2$\pm$0.4  &   2.702 \\ 
022150+1327 &   010803 &   3483 &   79  &   $-$0.2$\pm$0.2  &   2.376 \\ 
023430+2438 &   010125 &   4138 &   11  &   $-$0.1$\pm$0.4  &   2.196 \\ 
034857+1255 &   010125 &   4847 &   $<5$&   $\cdots$        &   3.296 \\ 
091848+2117 &   010218 &   2160 &   154 &   $-$0.6$\pm$0.1  &   2.274 \\ 
095504+1705 &   010129 &   4132 &   62  &   0.4$\pm$0.2     &   2.025 \\ 
102724+1219 &   001016 &   3249 &   35  &   0.2$\pm$0.2     &   2.057 \\ 
105144+3539 &   010213 &   4436 &   473 &   $-$0.2$\pm$0.1  &   2.105 \\ 
125807+2329 &   010730 &   3140 &   $<5$&   $\cdots$        &   2.066 \\ 
130005+1632 &   010730 &    887 &   127 &   0.2$\pm$0.1     &   2.199 \\ 
130700+2338 &   010527 &   3500 &   7   &   $\cdots$        &   3.342 \\ 
140251+2631 &   010704 &   1730 &   340 &   $-$0.6$\pm$0.1  &   2.114 \\ 
145331+1353 &   010601 &   3156 &   $<5$&   $\cdots$        &   2.298 \\ 
150113+2329 &   010726 &   3368 &   58  &   $-$0.3$\pm$0.2  &   2.412 \\ 
151653+1900 &   010427 &   1132 &   23  &   0.1$\pm$0.3     &   2.121 \\ 
163700+2221 &   010613 &   4371 &   163 &   $-$0.2$\pm$0.1  &   2.095 \\ 
165939+1834 &   010713 &   2406 &   24  &   0.6$\pm$0.3     &   2.167 \\ 
222554+1958 &   001002 &   3934 &   80  &   0.5$\pm$0.2     &   2.148 \\ 
234449+1221 &   010116 &   1937 &   244 &   $-$0.4$\pm$0.1  &   2.073 \\ 
\enddata
\tablenotetext{a} {Net Chandra broad band counts: 0.3$-$8.0 keV. Flux
conversion factor: 1 ct s$^{-1}$ $\to$ 1.137 $\mu$Jy at 1 keV, assuming \alpe=1.0, \nh=3
$\times 10^{20}$ cm$^{-2}$. }
\tablenotetext{b}{HR = hardness ratio: (H-S)/(H+S), Soft (S): 0.5$-$2.5 keV, Hard (H): 2.5$-$8.0 keV}
\end{deluxetable}


\clearpage

\clearpage
\end{document}